\documentclass[twocolumn,prb,showpacs,preprintnumbers,amsmath,amssymb,floatfix]{revtex4}
\usepackage{graphicx}
\usepackage{textcomp}

\begin{document}

\title{Structural properties and magnetoresistance of La$_{1.952}$Sr$_{0.048}$CuO$_4$ thin films}

\author{I Zaytseva$^1$, R Minikayev$^1$, E Dobro\v{c}ka$^2$, M \v{S}pankova$^2$, N Bruyant$^3$ and Marta Z Cieplak$^1$}

\affiliation{$^1$Institute of Physics, Polish Academy of Sciences, 02 668 Warsaw, Poland}

\affiliation{$^2$Institute of Electrical Engineering SAV, D\'{u}bravsk\'{a} cesta 9, 84104 Bratislava, Slovakia}

\affiliation{$^3$CNRS - LNCMI 143, avenue de Rangueil, 31400 Toulouse, France}

\begin{abstract}
{The evolution of the structural and transport properties of underdoped La$_{1.952}$Sr$_{0.048}$CuO$_4$ thin films under compressive epitaxial strain has been studied. The films of different thicknesses $d$ (from 26 nm to 120 nm) were deposited using an insulating target. The onset of superconductivity in the films is observed at temperatures as high as 26 K, while small residual resistance persists at low temperatures, indicating that superconductivity is inhomogeneous. The resistance measured under perpendicular magnetic field saturates below about 0.65 K, suggesting a possible existence of nonconventional metallic state. The magnetic-field-tuned superconductor-insulator transition is observed at magnetic field of about 32 T.}
\end{abstract}


\maketitle

\section {Introduction}

The properties of disordered or inhomogeneous superconducting systems are still not fully understood. Experiments show \cite{Gantmakher} that a superconductor to insulator transition (SIT) may be induced in thin superconducting films by decreasing of the film thickness, what enhances the disorder \cite{Markovic}, or by increasing the external magnetic field \cite{SIT}. In many such systems the scaling analysis of the resistance in the vicinity of the SIT follows the predictions for the dirty boson model in the vicinity of quantum critical point \cite{Fisher,FisherGrin}. On the other hand, the SIT is observed also in strongly inhomogeneous systems, such as granular superconducting films or Josephson junction arrays, which may be understood as a set of superconducting islands immersed in the metallic matrix \cite{Jaeger,Zant,Han,Eley}. In this case the SIT depends sensitively on the competition between the two energies: energy of Josephson coupling $E_J$ (that corresponds to the coupling between superconducting islands, allowing Cooper pair transport) and charging energy $E_C$, which, in turn, depends on two other parameters: distance between superconducting areas and the resistance of non-superconducting matrix \cite{Zant,Eley,Han}. Interestingly, some of these inhomogeneous systems appear to be metallic at low temperatures, but with unusually high resistance \cite{quantummetal}, inconsistent with the conventional theory of metals \cite{Abrahams}. It has been suggested recently that the island systems may be a "failed superconductor", in which current is carried by bosonic fluctuations, but the system fails to condense in the $T=0$ limit \cite{Kapitulnik2019}. The experimental study of superconducting films with tunable inhomogeneity should provide valuable insight into the nature of ground states in such systems.

In this work we tune the inhomogeneity of the superconducting films by utilizing the strain introduced by the lattice mismatch between the substrate and the film. It has been shown in the past that such strain is a very efficient method to modify superconductivity in thin films. In case of La$_{2-x}$Sr$_x$CuO$_4$ (LSCO) films the superconducting transition temperature, $T_c$, may be either suppressed or enhanced by the tensile or compressive strain, respectively, applied in the CuO$_2$ plane \cite{Sato,Locquet,Bozovic}. As we have shown previously \cite{Zaytseva2012}, the compressive in-plane strain may even induce superconductivity in the films deposited from non-superconducting LSCO target with $x=0.048$. In the present study we investigate structural and magneto-transport properties of La$_{1.952}$Sr$_{0.048}$CuO$_4$ thin films of different thicknesses, with a various epitaxial compressive built-in strain. At low $T$ the $T$-dependence of the resistance shows that the films behave as a system of irregular array of superconducting islands immersed in non-superconducting matrix. The transition to the zero-resistance superconducting state is not complete down to $T = 2$K, what may indicate that the value of $E_J$ is small due to the large distance between the superconducting island, or due to the strong inhomogeneity of charge carrier density \cite{Lin,Carbillet}.
The SIT is observed in the high transverse magnetic field in the film with thickness of 35 nm, with $T$-independent isotherm crossing at the magnetic field of about 32 T, and a finite-size scaling of resistance in the vicinity of the $B_c$. At the lowest temperatures the resistance appears to saturate, what may indicate the presence of anomalous metallic state.


\section {Experimental details}

Epitaxial La$_{1.952}$Sr$_{0.048}$CuO$_4$ (LSCO) thin films were deposited from stoichiometric ceramic target by a pulsed laser deposition (PLD) using Nd:YAG laser ($\lambda$ = 266 nm), with a repetition rate of 1 Hz and a pulse energy density 1.2 J/cm$^2$ at the target surface. The target with the Sr content  $x = 0.048$ is not a bulk superconductor. Films were grown on SrLaAlO$_4$ (SLAO) substrates of the area 5x5 mm$^2$. During deposition the substrates were held at temperature of 760$^{\circ}$C in the oxygen atmosphere of 300 mTorr. After deposition, the O$_2$ pressure in the chamber was increased to 500 Torr, and the films were slowly cooled down to room temperature with a rate of 3 K per minute.
The films were studied using X-ray diffraction with help of a laboratory X'Pert Pro MPD diffractometer. The out-of-plane lattice parameters $c$ were determined using XRD techniques for a series of thin films with thickness $d$ ranging between 26 nm and 120 nm. The reciprocal space maps of the films were measured in high-resolution mode on Bruker D8 DISCOVER diffractometer with rotating Cu anode operating at 12 kW (40 kV/ 300 mA). Superconducting transition temperature and magnetoresistance were measured on photolitographically patterned films using standard four-probe method in a Quantum Design PPMS (Physical Properties Measurement System) at $T \geq$ 2 K and in fields up to 9 T. In addition, some magnetotransport measurements were carried out at the Toulouse LNCMI high field facility, in pulsed high magnetic field up to 50 T, and in the temperature range 0.4 K $< T <$ 25K ($H \parallel c$, $I \parallel ab$).

\section {Results and Discussion}

\subsection{Structural properties}

The $a$-axis (in-plane) lattice parameter of SLAO substrates is equal to 3.757 {\AA}. This value is less than the in-plane lattice parameters of LSCO target with $x = 0.048$, which are equal to 3.806 {\AA} and 3.784 {\AA} for $a$ and $b$ lattice parameters, respectively.
Therefore, it may be expected that films will be compressed in-plane and expanded out-of-plane at the beginning of the deposition process.
In order to quantify the strain induced by the lattice mismatch we define the following strain parameters, $\varepsilon_l = (l_{film}/l_{bulk}-1)\ast 100\%$, where $l$ is the corresponding lattice parameter value.

\begin{figure}
\centering
\includegraphics[width=8.5cm]{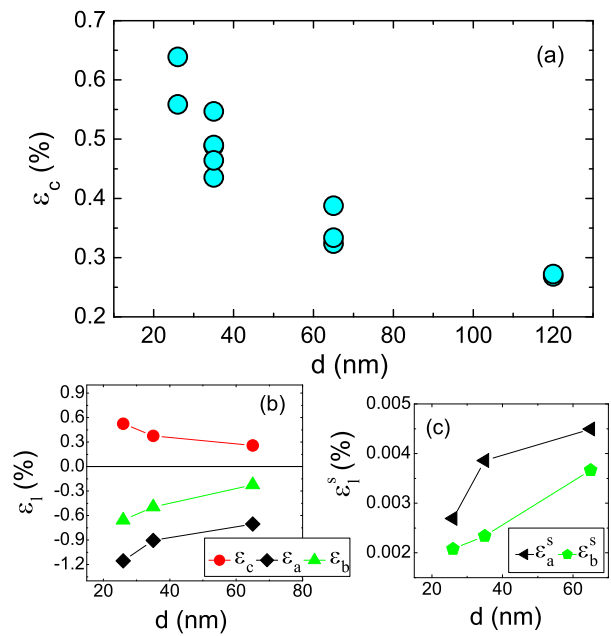}
\caption {(a) Dependence of $\varepsilon_c$ on film thickness $d$ for a series of films. (b) Strain parameters $\varepsilon_a$, $\varepsilon_b$ and $\varepsilon_c$ versus $d$ for three selected films. (c) The relative change of in-plane film lattice parameters with respect to the substrate parameters versus $d$. The size of the errors is less than the size of the symbols.}\label{fig1}
\end{figure}

Figure \ref{fig1}(a) shows the dependence of out-of-plane strain parameter $\varepsilon_c$ on the films thickness, $d$, for a large group of films. It is evident that the $\varepsilon_c$ is quite scattered for various films. Nevertheless, it is obvious that $\varepsilon_c$ is the largest for the thinnest film, and decreases as the film thickness increases. Figure \ref{fig1}(b) compares the $d$-dependence of in-plane strains ($\varepsilon_a$ and $\varepsilon_b$) and the out-of-plane strain $\varepsilon_c$ for three selected films with thicknesses 26 nm, 35 nm and 65 nm. In order to evaluate the in-plane lattice parameters two symmetric 006 and 0010 diffraction peaks and four asymmetric 1011 peaks at the azimuthal angles 0$^{\circ}$, 90$^{\circ}$, 180$^{\circ}$ and 270$^{\circ}$ were measured. The parameters $a$ and $b$ were determined separately using the the azimuthal orientation $0^{\circ} - 180^{\circ}$ and $90^{\circ} - 270^{\circ}$, respectively.

Fig. \ref{fig1}(b) confirms that the in-plane compression of $a$ and $b$ parameters is accompanied by the expansion of the $c$ lattice parameter. It also confirms that the largest value of $\varepsilon$ is for the thinnest films, consistent with the expectation that the strain induced by lattice mismatch is the largest at the beginning of the film deposition. As the thickness of the film increases, the atomic layers grow less compressed, and the film lattice parameters tend to the parameters of the bulk material. Finally, in Fig.\ref{fig1}(c) we show the $d$-dependence of the relative change of film parameters with respect to the parameters of the substrate, $\varepsilon^s_l = (l_{film}/l_{substrate}-1)\ast 100\%$. It is clear that in-plane parameters for all films are expanded in comparison with substrate in-plane parameters, and the expansion grows as film thickness increases. Nevertheless, the value of $\varepsilon^s_l$ remains very small in comparison to $\varepsilon_l$, confirming reasonably good matching between the lattice parameters of the substrate and the parameters of the strained films.

\begin{figure}
\centering
\includegraphics[width=8.5cm]{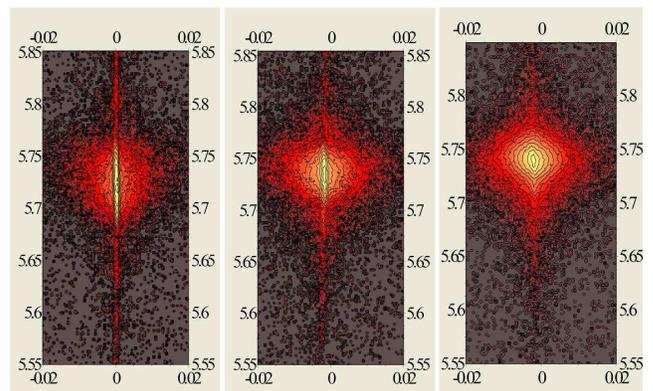}
\caption {RSM of layer maxima 006 for films with thickness (from left to right): 26 nm, 35 nm, 65 nm. RSM are presented with the coordinates $h$ and $l$ having the units $1h = 1/d_{100}$, and $1l = 1/d_{001}$, respectively. $d_{hkl}$ is the corresponding interplanar spacing of the substrate lattice.}\label{fig2}
\end{figure}

These results may be verified by measurements of the reciprocal space maps (RSM). In Fig.\ref{fig2} we show the RSM of the same three films, for which strain is shown in Fig.\ref{fig1}(b). Symmetric diffractions 006 are used for reciprocal space mapping, but only the layer peaks are mapped, without the substrate ones. The maps were recorded in the azimuthal direction 0$^{\circ} - 180^{\circ}$. Reciprocal space maps are presented with the coordinates $h$ and $l$ having the units $1h = 1/d_{100}$, and $1l = 1/d_{001}$, respectively. $d_{hkl}$ is the corresponding interplanar spacing of the substrate lattice. It is the advantage of this representation that the coordinates acquire an integer value at the diffraction spots of the reference substrate crystal. The maps indicate that the thinnest film (left on Fig.\ref{fig2}) is an undistorted thin film with almost constant lattice parameters across the whole film thickness. For thicker films the width of the layer maxima increases with thickness. This indicates the presence of the diffuse scattering due to distorted material, presumably resulting from misfit dislocations, producing the increase of mosaicity for thicker films.

\subsection{Temperature dependence of resistance}

\begin{figure}
\centering
\includegraphics[width=8.5cm]{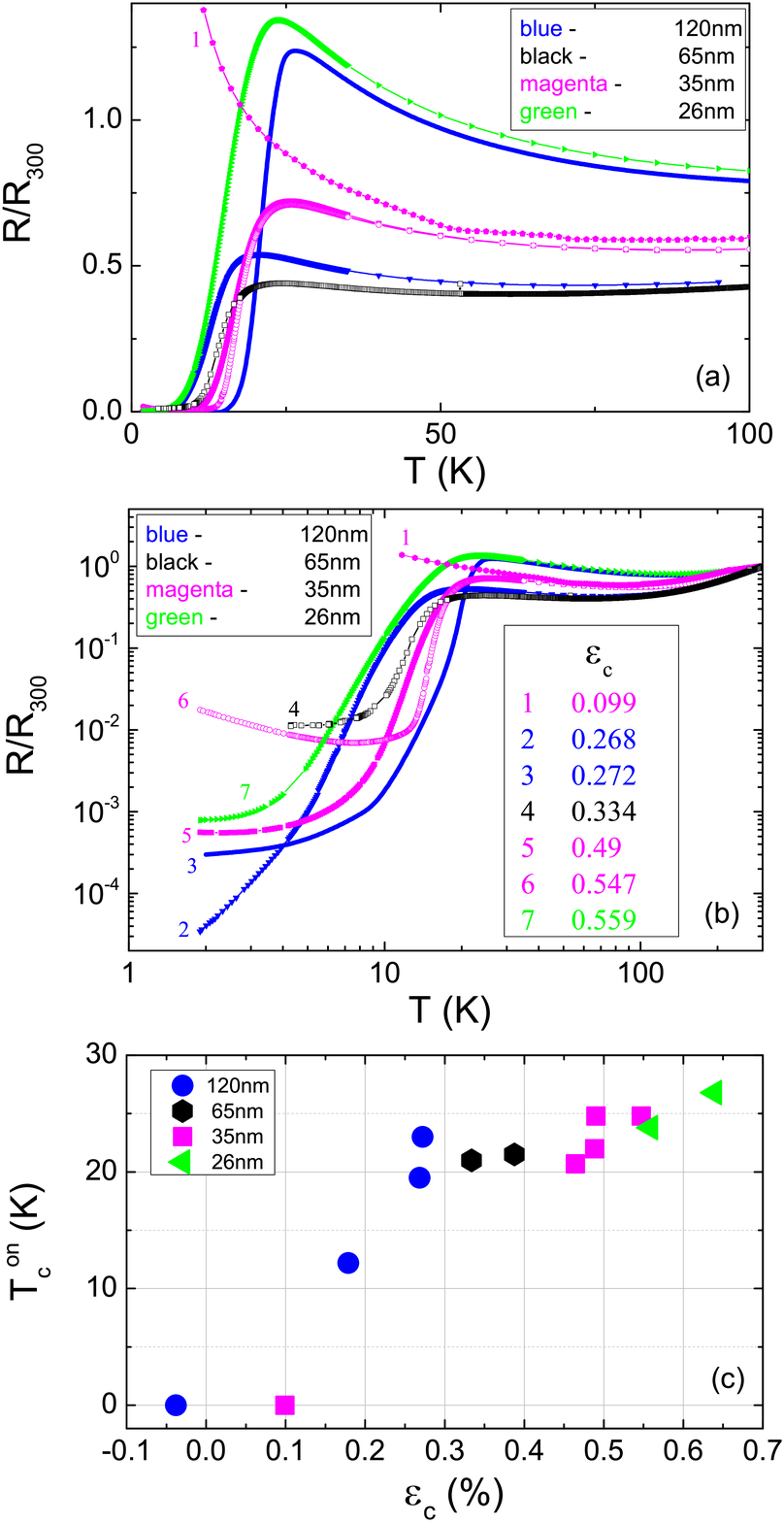}
\caption {$R/R_{300}$ vs $T$ on a linear scale (a) and on a log-log scale (b) for films with different $d$: 26 nm (green points), 35 nm (magenta), 65 nm (black), and 120 nm (blue), and with different $\varepsilon_c$, as indicated in (b) by labels 1 to 7. (c) $T^{on}_c$ versus $\varepsilon_c$ for films with different $d$. The size of errors is comparable to the size of symbols.}\label{fig3}
\end{figure}

The temperature dependence of resistance, normalized to room-temperature resistance, $R/R_{300}$, for several films with different thickness and different strain is shown in Fig. \ref{fig3}(a), and, on a double logarithmic scale, in Fig. \ref{fig3}(b). Except for one film with small $\varepsilon_c = 0.099$ (labeled 1), all other films with larger $\varepsilon_c$ show dramatic decrease of $R/R_{300}$ at low temperatures, suggesting the possible onset of superconductivity induced by in-plane compressive strain, as described previously in a preliminary report \cite{Zaytseva2012}. Several features are evident. First, samples with the same $d$ may show substantially different $T$-dependence of $R/R_{300}$ [Fig. \ref{fig3}(a)], indicating that, in addition to thickness, the strain strongly affects the resistance. The other feature, displayed in Fig. \ref{fig3}(b), is that the resistance does not reach zero as $T$ is lowered. Instead, all films show residual resistance, $R_{res}$, at $T$ = 2 K.

For further analysis, we define the onset of superconductivity, $T^{on}_c$, as the temperature, at which the resistance starts to deviate from the normal state resistance. Fig. \ref{fig3}(c) shows the dependence of the $T^{on}_c$ on strain $\varepsilon_c$ for films with different $d$. We observe reasonably good correlation between $d$, $\varepsilon_c$, and the $T^{on}_c$: as the film thickness increases, the strain relaxation occurs, and superconductivity weakens. Note that correlation between $d$ and $T^{on}_c$ is not strict, because in some thinner films the strain relaxation occurs, as seen in case of film 1 ($d = 35$ nm, $\varepsilon_c$ = 0.099), in which no traces of superconductivity are visible. On the other hand, the correlation between the $\varepsilon_c$ and $T^{on}_c$ appears to hold well. Interestingly, films with large strain, $\varepsilon_c > 0.25$, show relatively weak decrease of the $T^{on}_c$ with decreasing $\varepsilon_c$, from 27 K down to about 20 K, while more rapid decrease occurs for $\varepsilon_c < 0.25$. Thus, it seems that $d = 120$ nm is a limiting film thickness, above which rapid strain relaxation destroys superconductivity and the films become insulating, just as a bulk, underdoped target, with the Sr content $x$ = 0.048.

While we see good correlation between $\varepsilon_c$ and $T^{on}_c$, the behavior of resistance on the decrease of $T$ below $T^{on}_c$ is more complicated. Instead of zero-resistive state, the resistance of these films either flattens out at finite value, or, after reaching some minimum, it increases with further lowering of $T$. For example, while several films with $d$ = 35 nm exhibit similar $T^{on}_c$ [see Fig. \ref{fig3}(c)], the film with highest strain in this group (film 6) shows sharpest decrease of the resistance just below $T^{on}_c$, followed by a minimum at 10 K, and semiconducting-like increase of the resistance on further decrease of $T$ [Fig. \ref{fig3}(b)]. As a result, film 6 shows larger value of $R_{res}$ at 2 K than the film 5, in which strain is smaller. Such lack of correlation between the $R_{res}$ and $\varepsilon_c$ is also visible in case of two thick films, 2 and 3, which show almost identical (partially relieved) strain, but $R_{res}$ in these films differs by an order of magnitude.

The existence of residual resistance suggests that superconductivity in these strained films is inhomogeneous, so that no global phase coherence is reached, at least not down to 2 K. Inhomogeneous superconductivity may have many possible origins, including structural, chemical or charge density inhomogeneity. Structural origin has been reported, for example, in case of quench-deposited ultrathin amorphous Bi films, which display thickness variation of about 13\% of total film thickness \cite{Lin}. On the other hand, charge density inhomogeneity, not directly related to the topography of the sample surface, has been observed in ultrathin NbN films, suggesting that relation between the structural disorder and the charge inhomogeneity may be quite complex \cite{Carbillet}.

In case of LSCO with Sr content in the range $0.04 < x <0.16$ nuclear quadruple resonance experiments on bulk crystals and ceramic material uncovered charge density inhomogeneities of unknown origin on small length scales of 6-10 nm, while other probes (electron micro-probe and X-ray diffraction) show uniform charge distribution on larger length scales \cite{Singer}. Therefore, we cannot exclude the possibility that similar small length scale inhomogeneity occurs in our films. However, close correlation between the $\varepsilon_c$ and $T^{on}_c$ suggests that the principal origin in the present case is the heteroepitaxial growth of strained LSCO films on SLAO substrates. Such growth usually leads to the nucleation of misfit dislocations, which produce nonuniform strain distribution within the film. This, in turn, most likely leads to nucleation of superconductivity in spatially limited areas of the film, in which the strain is the strongest. In this case the system resembles granular superconductor, and it may be modeled as an disordered array of superconducting islands embedded in a metallic, non-superconducting matrix. This type of model has been used recently to discuss the experiment on an array of Nb dots deposited on a gold substrate, in which the Josephson coupling $E_J$ and the charging energy $E_C$ depend on the thickness of superconducting islands (i.e., Nb dots), and on distances between them \cite{Eley}.

Considering such model of superconducting islands we may explain the difference in the correlation between strain and the two parameters measured in our experiment, the $T^{on}_c$ and the $R_{res}$. The $T^{on}_c$ marks the temperature, at which on cooling of the very thin film the superconductivity first nucleates inside of the "islands", which in the present case are the areas of the film with the largest strain. Therefore, the $T^{on}_c$ is directly dependent on the magnitude of strain, which decreases with the increasing film thickness. On the other hand, the behavior of the resistance below $T^{on}_c$ depends not only on strain, but also on the thickness of the islands, and on the coupling between them, which, in turn, depends on the resistivity of the metallic matrix. In a very thin film the highly strained areas are limited in thickness, and the resistivity of the metallic matrix is increasing with lowering of $T$ due to carrier localization. Thus, after a rapid drop of resistance just below the $T^{on}_c$, we observe either a saturation of resistance at relatively large value of $R_{res}$, or even an increase of resistance on lowering of $T$, as seen in case of film 6. As the film thickness increases, the localized superconducting order within highly strained areas becomes more robust, and the coupling between islands becomes stronger, because the resistivity of the metallic matrix decreases. This evolution leads to a decrease of the $R_{res}$ in thicker films, although it may not be enough to achieve the global phase coherence.


\subsection{Magnetoresistance}

\begin{figure}
\centering
\includegraphics[width= 7cm]{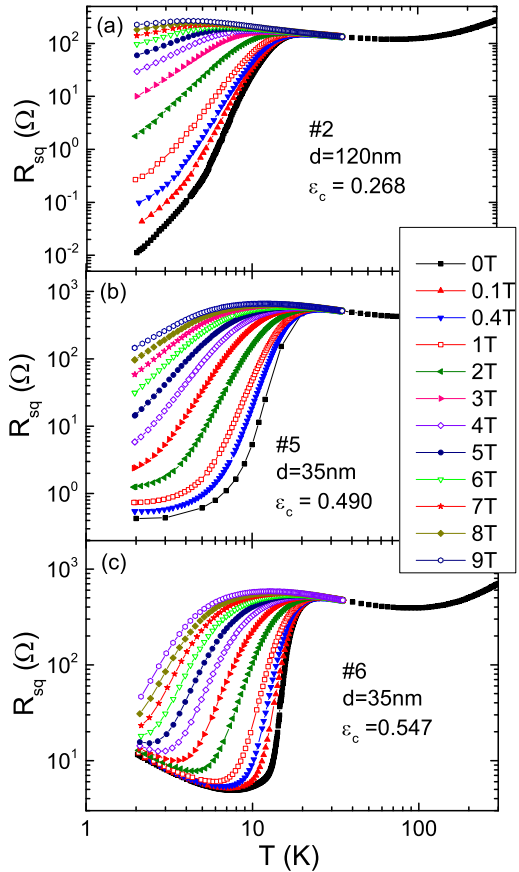}
\caption {$R_{sq}$ vs $T$ on a log-log scale, for $B$ in the range 0 to 9 T, for three films: film 2 with $d$ = 120 nm and $\varepsilon_c$ = 0.268 (a), film 5 with $d$ = 35 nm and $\varepsilon_c$ = 0.49 (b), and film 6 with $d$ = 35 nm and $\varepsilon_c$ = 0.547 (c).}\label{fig4}
\end{figure}

The $T$-dependence of the resistance per square, $R_{sq}$, measured in the presence of perpendicular magnetic field $B$, is presented on a double logarithmic scale in Fig. \ref{fig4} for three films, one with $d = 120$ nm and $\varepsilon_c$ = 0.268 (\ref{fig4}a), and two films with $d = 35$ nm but with different strain, film 5 with $\varepsilon_c$ = 0.49 (\ref{fig4}b) and film 6 with $\varepsilon_c$ = 0.547 (\ref{fig4}c). We observe that the suppression of superconductivity by the magnetic field becomes less effective as the strain grows, that is, on going from (\ref{fig4}a) to (\ref{fig4}c).

It is also interesting to see that the weak magnetic field has a different effect on the magnitude of residual resistance in these films. In the case of thickest film with $d = 120$ nm the $R_{res}$ increases by 3 orders of magnitude when the magnetic field increases from 0 to 4 T, what is the result of the usual broadening of superconducting transition, caused by the decrease of the activation energy for vortex pinning with the increasing magnetic field. On the other hand, in case of the films with $d = 35$ nm the $R_{res}$ either does not change at all in the same field range (film 6) or it changes only weakly (film 5); moreover, the $T$-dependence of the resistance at the lowest $T$ remains insulating-like in the first case, and metallic in the second case. This suggests that at weak magnetic fields the principal contribution to the $R_{res}$ originates from the resistance of the normal, metallic regions of the sample. With increasing magnetic field the resistance minimum, which is seen in case of film 6, shifts towards lower temperatures. Faster increase of $R_{res}$ starts only after the minimum disappears, when $B$ exceeds about 4 T. This is when the broadening of superconducting transition due to vortex unpinning from locally superconducting areas starts to contribute to the $R_{res}$. We may conclude that the evolution of resistance with magnetic field supports the scenario of inhomogeneous superconductivity in the strained LSCO films.

\begin{figure}
\centering
\includegraphics[width= 8.5cm]{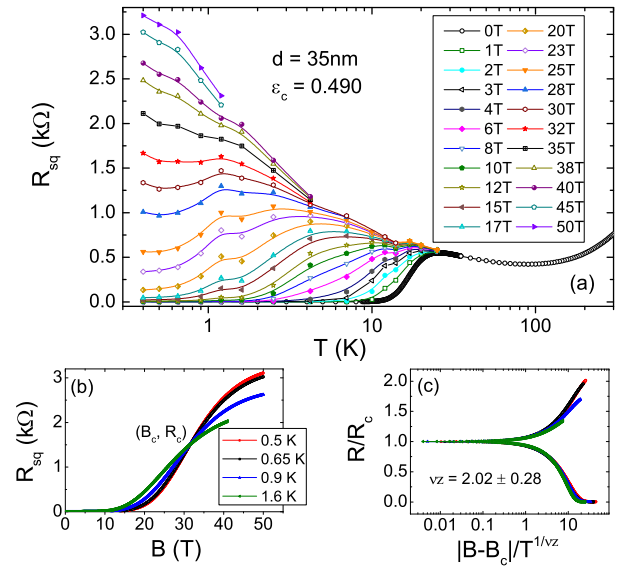}
\caption {(a) $R_{sq}$ vs $T$ for various $B$ for film with $d$ = 35  nm. (b) The T-independent crossing point of isotherms at $B$ = 31.79 T and $R_c =$ 15.44 k$\Omega$. (c) Resistance as a function of scaling variable, $|B-B_c|T^{-1/\nu z}$.}\label{fig5}
\end{figure}

Further increase of the magnetic field induces a superconductor-insulator transition. This is illustrated in Fig. \ref{fig5}(a), which shows the $T$-dependence of the $R_{sq}$ for the film 5 ($d = 35$ nm, $\varepsilon_c$ = 0.49), measured in high pulsed magnetic fields up to 50 T. The data were extracted from the magnetoresistance measurements at fixed $T$. There is some scatter of the data, particularly for $T$ = 0.4 K and $T$ = 1.2 K, probably due to temperature instability during the measurements. Nevertheless, the curves show clearly a gradual transition from the region with $dR_{sq} /dT > 0$ at low magnetic fields to the region with $dR_{sq} /dT < 0$ at high magnetic fields. The crossover between these two regions occurs in the vicinity of $B = 32$ T. Interestingly, the resistance shows a tendency to saturate at the lowest temperatures ($T < 0.65$ K). We also note that this saturation seems to persists up to highest fields (45 T and 50 T).

Replotting the data versus $B$ in Fig. \ref{fig5}(b) we find a point of isotherm crossing at $B_c =$ 31.79 T and $R_c =$ 15.44 k$\Omega$ for the temperature range 0.5 K $< T <$ 1.6 K (except for slight deviations at temperatures 0.4 K and 1.2 K, as mentioned above). Such crossing is usually identified as a quantum critical point, at which the SIT takes place \cite{Gantmakher,SIT}. Scaling analysis has been performed in the vicinity of the $B_c$. Fig. \ref{fig5}(c) shows the data plotted versus scaling variable according to equation, $R/R_c = f(|B-B_c|T^{-1/\nu z})$, where $z$ and $ \nu $ are dynamic and correlation critical exponents, respectively.
The scaling exponents $\nu z$ are found to be 2.02 $\pm$ 0.28. The value of $\nu z >$ 1 corresponds to the exponent predicted in the framework of a dirty boson picture ($\nu >$ 1, assuming $z$ = 1) and this model describes the quantum transition at $T$ = 0 in a 2D disordered system \cite{Fisher,FisherGrin}.

It is worth noting that we have not observed two crossing points in the same film, as reported by Shi \textit{et al.} \cite{Shi} for LSCO films with $x$ = 0.07. In that case two different crossing points have been found, one at high-$T$ (low $B$) range, and another at low-$T$ (high $B$) range, with critical exponents $\nu z$ equal to 0.737 and 1.15, respectively. These crossing points have been attributed to two transitions, one from pinned vortex solid to vortex glass, and another from vortex glass to insulator, respectively. It is the first transition, at temperatures just below $T^{on}_c$, which apparently does not occur in the case of strained film 5, measured in the present experiment. It is possible that the pinned vortex solid state is absent in present case, because of inhomogeneous nature of superconductivity in our films. Another possibility is that it cannot be observed because the resistance contains contributions from two different regions of the film, superconducting islands, and the normal metallic matrix. What is interesting is that despite such two contributions we still observe a second transition, at low $T$ (high $B$) range.

Finally, we note that the saturation of the resistance at very low temperatures (below about 0.65 K), which seems to persist up to high fields [Fig.\ref{fig5}(a)], deserves further study. Similar phenomenon, which has been reported in many other superconducting systems with SIT \cite{Kapitulnik2019}, is attributed to the existence of unconventional metallic state. The nature of this state, which so far is not well understood,  is a subject of many studies and discussions.

\section{Conclusions}

We report on the resistance and magnetoresistance measurements in La$_{1.952}$Sr$_{0.048}$CuO$_4$ thin films, in which in-plane compressive strain is induced by lattice parameters mismatch to the substrate. The strain induces superconductivity in thin films deposited from non-superconducting target. The evolution of resistance with temperature and magnetic field supports the scenario of inhomogeneous superconductivity, which resembles a disordered array of superconducting islands immersed in a nonsuperconducting matrix. The magnetic-field-tuned superconductor-insulator transition is observed at magnetic field of about 32 T, while the saturation of resistance below about 0.65 K may indicate the existence of nonconventional metallic state.

\section*{Acknowledgments}
We would like to thank M. Berkowski and M. G\l{}owacki for experimental support. This research was partially performed in the NanoFun laboratories co-financed by the ERDF Project NanoFun POIG.02.02.00-00-025/09. We acknowledge the support of LNCMI-CNRS a member of the European Magnetic Field Laboratory (EMFL).

\end{document}